\documentclass[10pt,twocolumn,letterpaper]{article}
\pdfoutput=1
\usepackage{cvpr}       
\usepackage{graphicx}
\usepackage{amsmath}
\usepackage{amssymb}
\usepackage{multirow}
\usepackage{booktabs}
\usepackage{url}
\usepackage[accsupp]{axessibility}

\usepackage[pagebackref,breaklinks,colorlinks]{hyperref}

\newcommand\blfootnote[1]{%
	\begingroup
	\renewcommand\thefootnote{}\footnote{#1}%
	\addtocounter{footnote}{-1}%
	\endgroup
}
\usepackage{tikz}

\usepackage[capitalize]{cleveref}
\crefname{section}{Sec.}{Secs.}
\Crefname{section}{Section}{Sections}
\Crefname{table}{Table}{Tables}
\crefname{table}{Tab.}{Tabs.}

\begin{document}
\title{
\textcolor{red}{EAGLE}: Cont\textcolor{red}{e}xtual Point Cloud Generation via \textcolor{red}{A}daptive \\ Continuous Normalizin\textcolor{red}{g} F\textcolor{red}{l}ow with S\textcolor{red}{e}lf-Attention}

\author{
    Linhao Wang $^{1}$, 
    Qichang Zhang $^{2}$, 
    Yifan Yang $^{3}$, 
    Ye Su $^{1}$,
    Hao Wang $^{1,\dag}$
    \\
    $^{1}$ Shandong Normal University\\
    $^{2}$ University of Macau \\
    $^{3}$ Huazhong University of Science and Technology
}

\maketitle
\blfootnote{\dag Corresponding Author}
% \blfootnote{\ddag Project Leader}
% \blfootnote{This work has been submitted to the IEEE for possible publication. Copyright may be transferred without notice, after which this version may no longer be accessible.}

%%%%%%%%% ABSTRACT
\begin{abstract}
As 3D point clouds become the prevailing shape representation in computer vision, generating high-quality point clouds remains a challenging problem. Flow-based models have shown strong potential due to exact likelihood estimation and invertible mappings. However, existing flow-based methods for point clouds typically rely on point-wise feature extractors, which limits their ability to model long-range dependencies and global structural relationships among points. Inspired by the wide adoption of Transformers, we explored the complementary roles of self-attention mechanisms, CNN, and flow-based model. To this end, we propose EAGLE, a probabilistic generative model that integrates self-attention mechanisms with adaptive continuous normalizing flows. The self-attention module explicitly models pairwise dependencies among points, enabling effective capture of global contextual information. In addition, we introduce an adaptive bias correction mechanism within flow-based models, which dynamically adjusts to different input contexts and alleviates bias-drift issues. Extensive experiments on ShapeNet and ModelNet datasets demonstrate the effectiveness of our proposed method.
\end{abstract}

%%%%%%%%% BODY TEXT
\section{Introduction}
\label{sec:intro}
Point clouds have recently garnered significant interest due to their increasing popularity as a representation of 3D objects. Point clouds can represent geometric details more precisely than voxel grids or meshes while occupying less space. They also support emerging fields such as robotics, virtual/augmented reality, and autonomous driving. Numerous studies have made strides in advancing point cloud generation, achieving remarkable results \cite{mo2023dit,wang2024dpr, zhang2025enhancing,wang2024neighborhood, gadelha2018multiresolution,zamorski2018adversarial,achlioptas2018learning,wu2023sketch}.   Many of these approaches represent point cloud distributions as a fixed-dimensional matrix to simplify model processing. However, this fixed-dimensional representation limits these models to handling only a set number of points. When the actual number of points is lower, upsampling is required. On the other hand, downsampling must occur, often resulting in the loss of essential point cloud features. Furthermore, representing point clouds as fixed-dimension matrices is not well suited for large-scale point sets. Moreover, point clouds inherently exhibit geometric invariances, such as translation and rotation, which are often ignored by fixed-matrix representations. As a result, models must rely on additional parameters to learn these invariances, leading to reduced parameter efficiency.

Previous research has explored various generative models for point cloud generation. PointFlow \cite{yang2019pointflow} models the distributions of both shapes and points using continuous normalizing flows, enabling the generation of point clouds with an arbitrary number of points. Similarly, DPM \cite{luo2021diffusion} introduces a shape latent variable and models point distributions through normalizing flows \cite{dinh2016density,rezende2015variational} and diffusion processes \cite{ho2020denoising}, achieving promising results in point cloud generation.

Despite their success, existing flow-based point cloud generation methods primarily adopt point-wise architectures to extract features. Such designs inherently rely on independent feature aggregation, which limits their ability to explicitly model long-range dependencies and global contextual relationships among points. As a result, capturing complex global structures and inter-point correlations still remains challenging.

Inspired by the effectiveness of the Transformer architecture \cite{vaswani2017attention}, we explore the combination of self-attention mechanisms and flow-based model for point cloud generation. Self-attention models pairwise interactions among points, enabling the model to adaptively focus on informative features across the entire input points. However, deep networks can cause gradient issues (exploding or vanishing gradients) during backpropagation due to accumulation of gradients. To ensure stable and effective training, we adopt residual connections in the self-attention module. In addition, we also introduce residual modules before and after the self-attention layer, which preserve original point features while seamlessly integrating global contextual information, resulting in more stable feature representations throughout the network.

Furthermore, standard continuous normalizing flow models may suffer from bias drift during training, which can negatively affect generation performance. To address this issue, we propose Adaptive Continuous Normalizing Flows (A-CNF), which introduce an adaptive bias correction mechanism with learnable scaling factors for bias terms. This design dynamically adjusts to different input contexts and mitigates bias-drift issues.

Based on these insights, we propose EAGLE, a probabilistic model for contextual point cloud generation that integrates self-attention mechanisms with adaptive flows. Extensive experiments on ShapeNet and ModelNet datasets demonstrate that our approach more effectively captures global contextual features and achieves superior performance in high-quality point cloud generation.

The main contributions of our work are summarized as follows:
\begin{itemize}
    \item We propose EAGLE, a novel probabilistic framework for contextual point cloud generation, which integrates self-attention mechanism with adaptive continuous normalizing flows to effectively capture complex global dependencies and structural relationships among points.
    \item We introduce adaptive continuous normalizing flows with an adaptive bias correction mechanism, which dynamically adjusts the flow dynamics during training and effectively mitigates the bias drift issue.
    \item Extensive experiments on ShapeNet and ModelNet datasets demonstrate that EAGLE consistently outperforms state-of-the-art methods in terms of generation quality.
\end{itemize}

%-------------------------------------------------------------------------
\section{Related Work}
\paragraph{\textbf{Point Cloud Generation}} Point cloud data, which is fundamentally composed of 3D coordinates, is characterized by sparsity and unordered structure. Previous works have converted point cloud distributions into $N \times 3$  matrices, where \textit{N} is the fixed number of points, to facilitate processing. Achlioptas \textit{et al.} \cite{achlioptas2018learning} employ generative adversarial networks \cite{arjovsky2017wasserstein,gulrajani2017improved,goodfellow2014generative} for point clouds. Gadelha \textit{et al.} \cite{gadelha2018multiresolution} explore a tree-structured network via a variational auto-encoder \cite{kingma2013auto} to generate 3D point clouds. Zamorski \textit{et al.} \cite{zamorski2018adversarial} introduce adversarial autoencoders \cite{makhzani2015adversarial} for point cloud generation. However, a key limitation of these works is that they only generate a fixed number of points, overlooking permutation invariance. FoldingNet \cite{yang2018foldingnet} and AtlasNet \cite{groueix2018papier}  partially addressed this issue by converting 2D patches to 3D point clouds. Both methods allow for generating an arbitrary number of points while preserving permutation invariance. These methods rely on heuristic loss functions, such as the Chamfer Distance (CD) and Earth Mover’s Distance (EMD) \cite{fan2017point}, to calculate distances between point sets. However, EMD is computationally slow and its approximation can lead to biased gradients. On the other hand, CD fails to consider the point density distribution within the point cloud and is sensitive to noise and outliers.

To improve the representation of point clouds,  \cite{yang2019pointflow,luo2021diffusion} introduced probabilistic distribution frameworks where each point cloud is treated as sample data from a distribution. \cite{yang2019pointflow} models the distributions of both points and shapes via continuous normalizing flows, allowing for sampling an arbitrary number of points to represent the point cloud. This approach enables joint learning of distributions within both latent space and point space, generating high-quality point clouds while avoiding the limitations of heuristic loss functions. \cite{luo2021diffusion} similarly models distributions using normalizing flows , training a diffusion model directly on point cloud data. However, existing flow-based methods for point clouds typically
rely on point-wise feature extractors, making it difficult to model long-range dependencies and global structural relationships
among points. Inspired by the architecture of Transformer \cite{vaswani2017attention}, we use a self-attention mechanism to explicitly model pairwise dependencies among points,
enabling effective capture of global contextual information. This strengthens the model’s comprehension ability and effectively alleviates the mentioned drawbacks.
\paragraph{\textbf{Normalizing Flow}} Normalizing flow (NF) is a generative model that progressively transforms a simple distribution into a more complex data distribution through a series of invertible transformations \cite{rezende2015variational,dinh2016density}. Continuous normalizing flow (CNF) further extends this framework by using a sequence of continuous transformations described by ordinary differential equations \cite{chen2018neural,grathwohl2018ffjord}. Instead of constructing the flow through function composition, this approach formulates the flow as a continuous-time dynamic, allowing for smooth and continuous transformations from noise distribution to data distribution. Most research on normalizing flows has focused on image and simple data generation \cite{papamakarios2017masked,yao2023local}. However, works such as \cite{yang2019pointflow,postels2021go,kim2020softflow,klokov2020discrete,luo2021diffusion} have applied normalizing flows to point cloud generation tasks, achieving significant results. Our method builds upon this line of work, using adaptive continuous normalizing flows with an innovative improvement to the fundamental layer of CNFs.
\paragraph{\textbf{Attention Mechanisms for Point Cloud}}
Many works have explored the potential of attention mechanisms in point cloud tasks.
PointGrow \cite{sun2020pointgrow} introduces dedicated self-attention modules to capture long-range dependencies in the shape of point cloud objects. Point Transformers \cite{wu2022point, zhao2021point} adapt global attention to local attention to reduce memory usage and computational complexity.
PCT \cite{guo2021pct} employs global attention for point cloud processing. 
% TGN \cite{xu2023transformer} presents a Transformer-based framework, which performs progressive generation and refinement from feature space to coordinate space. 
Tiger \cite{ren2024tiger} aggregates global information using Transformers while employing CNNs to model local information. Contrary to previous methods, our approach combines the self-attention mechanism with adaptive flows to model pairwise interactions among points and capture global contextual information.

\section{Preliminaries}
In this section, we first provide an overview of the concept of flow-based models, followed by an introduction to the theory of continuous normalizing flows, variational autoencoder and the fundamental training framework.
\subsection{Overview of Flow-Based Models}
The flow-based model serves as the core framework, describing the transformation process of a data distribution \( p(x) \) through reversible mappings. The probability density transformation follows:
$p_{\theta}(x) = p_{\theta}(z) \left| \det \left( \frac{\partial x}{\partial z} \right) \right|^{-1},$ where \( x \in \mathbb{R}^m \) is a sample vector in the observation data space, representing the 3D point cloud we aim to generate, and \( z \in \mathbb{R}^n \) is a latent variable sampled from a high-dimensional distribution, capturing latent features of the point cloud. The mapping function \( f_{\theta}: z \to x \) represents the generative process, establishing the dependency between the generation and the inference processes.

Using the chain rule, the Jacobian of the transformation is given by: $
\frac{\partial x}{\partial z} = \frac{\partial f_{\theta}(z)}{\partial z}$. This Jacobian matrix \( J = \frac{\partial x}{\partial z} \) is critical in determining the relationship between input and output, indicating the sensitivity of the generated point cloud \( x \) to variations in the latent variable \( z \).

By computing the Jacobian matrix, we can analyze how small variations in input features influence the generated features, which facilitates model parameter adjustments during training. Moreover, the Jacobian is an essential tool for computing the gradients of the log-likelihood \cite{dinh2014nice}. This enables efficient optimization of the generative model using gradient-based methods. The transformation process allows complex data distributions to be mapped to simpler distributions (e.g., a multivariate 3D Gaussian), supporting efficient sample generation and density estimation.

\subsection{Continuous Normalizing Flow} 
Normalizing Flow \cite{rezende2015variational,papamakarios2021normalizing} aims to transform a simple known distribution (e.g., a 3D Gaussian) into a complex target distribution through a series of easily computable and reversible transformations. Let  $p_z(z)$  represent a simple latent distribution. This distribution is transformed into the target distribution  $p_x(x)$  through a series of reversible transformations  $f_1, f_2, \ldots, f_N$. The transformation can be expressed as
$x = f_N \circ f_{N-1} \circ \cdots \circ f_1(z)$. Here, $z$ denotes a latent variable sampled from a simple prior distribution
(e.g., a 3D Gaussian), and $x$ is obtained via a sequence of reversible
transformations. Let $z_0 = z$ and $z_k = f_k(z_{k-1})$ for $k=1,\dots,N$.
The inverse mapping is given by
$z_0 = f_1^{-1} \circ \cdots \circ f_N^{-1}(x)$. The probability density of the output variable $x$ is given by
\begin{align}
p_x(x) = p_z(z_0) \prod_{k=1}^{N} \left| \det \left( \frac{\partial f_k}{\partial z_{k-1}} \right) \right|^{-1},
\end{align}
where the density is accumulated through the change of variables at each layer. Taking the logarithm of both sides yields
\begin{align}
\log p_x(x) = \log p_z(z_0) - \sum_{k=1}^{N} \log \left| \det \left( \frac{\partial f_k}{\partial z_{k-1}} \right) \right|.
\end{align}
Continuous Normalizing Flow (CNF), however, extends the normalizing flow framework by modeling continuous-time transformations using neural networks and solving ordinary differential equations (ODEs) \cite{chen2018neural}, allowing smooth and continuous transformation of probability density over time. This transformation is represented by the continuous-time dynamics equation
\begin{align}
\frac{\partial z(t)}{\partial t}=f(z(t), t),
\end{align}
where $f$ is a neural network parameterized by time $t$ and state $z(t)$. The target distribution $p(x)$ is obtained by transforming the prior distribution  $p(z)$ . Starting from the initial state $ z(t_0)$  sampled from  $p(z)$ , the target state  $x$  is obtained through the continuous transformation
\begin{align}
x = z(t_0) + \int_{t_0}^{t_1} f(z(t), t) \, dt,
\end{align}
The inverse mapping can be calculated by
\begin{align}
z(t_0) = x + \int_{t_1}^{t_0} f(z(t), t) \, dt,
\end{align}
CNFs vary time $t$ from $t_0$ to $t_1$ and use a trace term to replace the Jacobian determinant in discrete normalizing flows, allowing for efficient computation. The log-probability density formula for continuous normalizing flows is described as
\begin{align}
\log p_x(x) = \log p_z(z(t_0)) - \int_{t_0}^{t_1} \operatorname{tr} \left( \frac{\partial f(z(t), t)}{\partial z(t)} \right) dt.
\end{align}
\subsection{Variational Autoencoder}
The Variational Autoencoder (VAE) \cite{kingma2013auto} aims to jointly minimize the reconstruction error between the input and output while encouraging the approximate posterior
$Q_\varphi(z \mid x)$ to be close to a prior distribution $P_\eta(z)$.
During training, the encoder $Q_\varphi(z|x)$ and the decoder $P_\alpha(x|z)$
are optimized by maximizing the evidence lower bound (ELBO) formulated as
\begin{align}
\log P_\alpha(x) &\geq E_{Q_\varphi(z|x)}\left[\log P_\alpha(x|z)\right] \notag \\
                 &- D_{KL}\left(Q_\varphi(z|x) \parallel P_\eta(z)\right) \notag\\
                 &= L_{\varphi,\alpha,\eta}(x).
                 \label{eq:ELBO}
\end{align}
% \begin{figure*}[ht]
%     \centering
%     \includegraphics[width=\textwidth]
%     {main_framework.pdf}
%     \caption{The visualization of the proposed method. (A) illustrates how the objective is computed during the training stage. (B) illustrates the sampling (generation) stage. The ConvBlock consists of 1D convolution layers, batchnorm layers and ReLU activation function. The latent block maps global point cloud features to the parameters of a latent distribution, from which latent variables are sampled via the reparameterization trick for subsequent generative modeling. }
%     \label{fig:main}
% \end{figure*}
\begin{figure*}[ht]
    \centering
    \includegraphics[width=\textwidth]
    {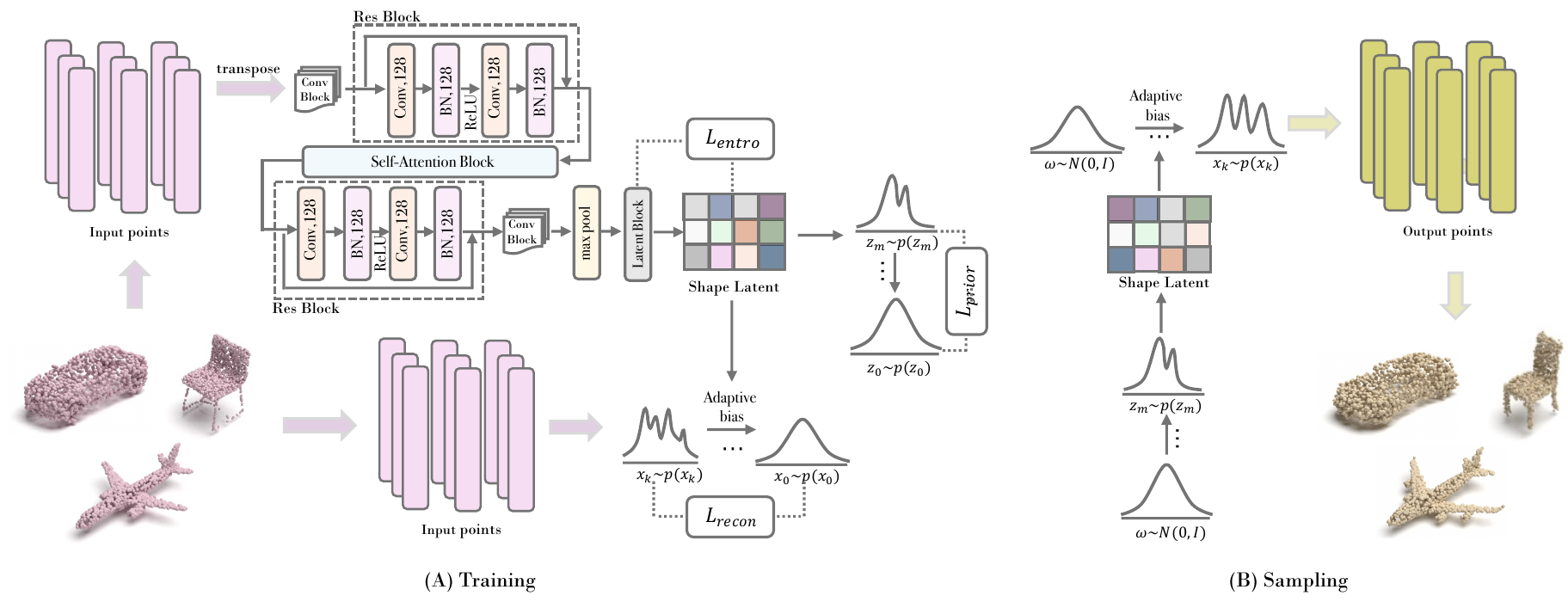}
    \caption{The visualization of the proposed method. (A) illustrates how the objective is computed during the training stage. (B) illustrates the sampling (generation) stage. The ConvBlock consists of 1D convolution layers, batchnorm layers and ReLU activation function. The latent block maps global point cloud features to the parameters of a latent distribution, from which latent variables are sampled via the reparameterization trick for subsequent generative modeling. }
    \label{fig:main}
\end{figure*}
\begin{figure}
    \centering
    \includegraphics[width=1\linewidth]{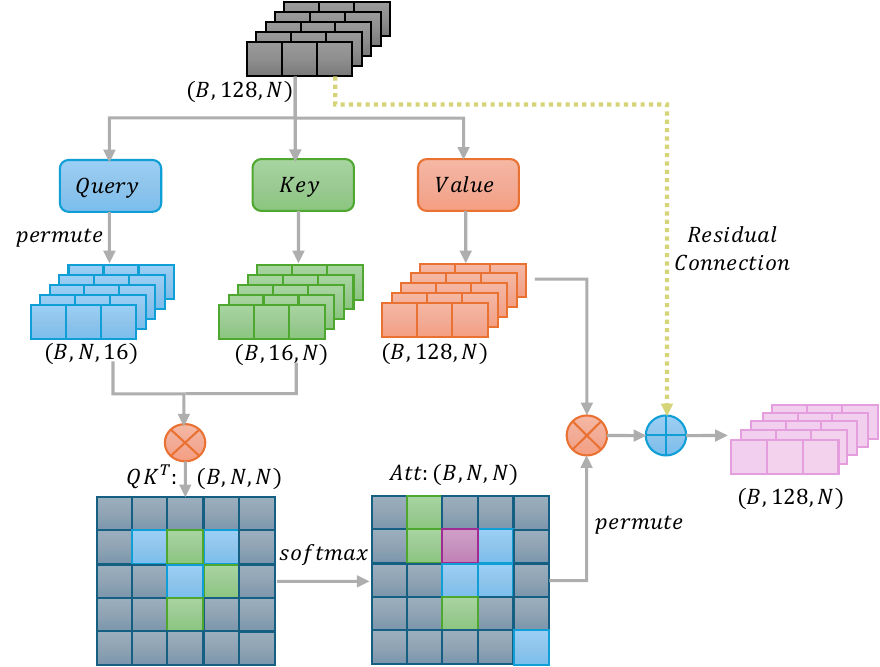}
    \caption{Visualization of the structure of our proposed Self-Attention block. }
    \label{fig:transformer}
\end{figure}
\subsection{Fundamental Training Framework}
\label{sec:ftf}
The most relevant work to ours is PointFlow \cite{yang2019pointflow}, which proposes a 3D point cloud generation model based on continuous normalizing flows. This model represents point clouds as a distribution of distributions, capturing both the distribution of shapes and the distribution of points within each shape. PointFlow formulates the reconstruction likelihood and the log probability of the prior distribution as Eq. \ref{eq:reconstruction} and \ref{eq:prior}.
\begin{align}
\log P_\alpha\left(x\middle| z\right) &= \log P\left(G_\alpha^{-1}\right) - \int_{t_0}^{t_1} \operatorname{tr}\left(\frac{\partial g_\alpha}{\partial y\left(t\right)}\right) dt,
\label{eq:reconstruction}
\end{align}

\begin{align}
\log P_\eta(z) = \log P\left(F_\eta^{-1}\right) - \int_{t_0}^{t_1} \operatorname{tr}\left(\frac{\partial f_\eta}{\partial \omega\left(t\right)}\right) dt.
\label{eq:prior}
\end{align}
The inverse of $F_\eta$ and $G_\alpha$ are each defined as
\begin{align}
F_\eta^{-1}(z): \omega\left(t_0\right) &= z -\int_{t_0}^{t_1} f_\eta\left(\omega\left(t\right), t\right)dt, \\ 
G_\alpha^{-1}(x,z): y\left(t_0\right) &= x - \int_{t_0}^{t_1} g_\alpha\left(y\left(t\right),z,t\right)dt.
\end{align}
While we follow the same two-level hierarchical training framework as PointFlow \cite{yang2019pointflow}, our key lies in the formulation of the conditional CNFs. Specifically, we introduce adaptive continuous normalizing flows (A-CNFs) by incorporating an adaptive bias correction mechanism into the base ODE layers. This design dynamically adjusts the bias magnitude under different latent contexts and effectively mitigates bias-drift issues.

\section{Method}
In this section, we dive into the details of our proposed model. The visualization of the proposed method is shown in Fig. \ref{fig:main}. We begin by discussing the context-aware encoder $Q_\varphi(z|x)$ equipped with residual connections \cite{he2016deep}, followed by adaptive continuous normalizing flows (A-CNFs), focusing on the improvements to the base ODE layers. Finally, we formulate the overall objective. Experimental results will be provided in the following section.
\subsection{Context-Aware Encoder with Self-Attention}
We employ a PointNet-style encoder \cite{qi2017pointnet} as the backbone of our point cloud feature extractor. Specifically, the point encoder takes $x \in \mathbb{R}^{B \times N \times 3}$ as input, where $B$ denotes the batch size and $N$ represents the number of points. Following \cite{qi2017pointnet,yang2019pointflow}, the input is first transposed to $\mathbb{R}^{B \times 3 \times N}$ and mapped to a 128-dimensional point-wise feature space via a 1D convolution with kernel size 1.

While PointNet aggregates global information through symmetric pooling operations, it fails to explicitly model interactions among individual points. To address this limitation and enhance global contextual modeling, we introduce a self-attention module that explicitly captures pairwise dependencies among points. As illustrated in Fig.~\ref{fig:transformer}, a single self-attention layer is inserted after the initial point embedding stage, where point-wise features are sufficiently expressive yet retain fine-grained geometric information. Specifically, the query and key projections reduce the channel dimension from 128 to 16, while the value projection preserves the original 128-dimensional feature space. The attention scores are calculated as 
\begin{align}
\varphi_{score}(Q,K)=\frac{QK^{T}}{\sqrt{d_k}},
\end{align}
followed by a softmax operation to obtain attention weights
\begin{align}
\alpha_{i j}=\frac{e^{\varphi_{score}(q_{i},k_{j})}}{\sum_{j}e^{\varphi_{score}(q_{i},k_{j'})}}. 
\end{align}
The output feature is then obtained as a weighted sum over all point features:
\begin{align}
\Lambda = \mathrm{softmax}\left(\frac{QK^{T}}{\sqrt{d_k}}\right)V,
\end{align}
where $d_k$ denotes the dimensionality of the key embeddings \cite{vaswani2017attention}. This formulation enables each point to attend to all other points, thus capturing long-range dependencies of points.

However, incorporating self-attention into deep architectures may introduce optimization difficulties due to vanishing gradients \cite{he2016deep}. To ensure stable training, we adopt residual connections in the self-attention module formulated as $y = \mathrm{ReLU}(F(x) + x)$. In addition, we place residual blocks both before and after the self-attention layer as shown in Fig. \ref{fig:main} (A). Each residual block consists of two 1D convolution layers with batch normalization and ReLU activation. This design not only enhances feature expressiveness before the self-attention module, but also stabilizes the output representations afterward. The overall structure can be summarized as
\begin{align}
\begin{array}{l}
x = \mathrm{SelfAttention}(\mathrm{ResidualBlock}_1(x)), \\
x = \mathrm{ResidualBlock}_2(x).
\end{array}
\end{align}
We follow the same approach as \cite{yang2019pointflow,qi2017pointnet} for subsequent processing of point clouds. Ablation studies further demonstrate that the integration of self-attention with residual blocks leads to consistent performance improvements.

\subsection{Adaptive Continuous Normalizing Flow}
Following \cite{yang2019pointflow}, we employ two continuous normalizing flows to model
the latent prior and the conditional data distribution, respectively. Specifically, the prior CNF defines an \emph{unconditional} flow for the latent variable,
while the decoder CNF defines a \emph{conditional} flow over the data space given the latent
shape \(z\), which serves as a compact representation of the input point cloud.
\begin{figure*}
    \centering
\includegraphics[width=\textwidth]
    {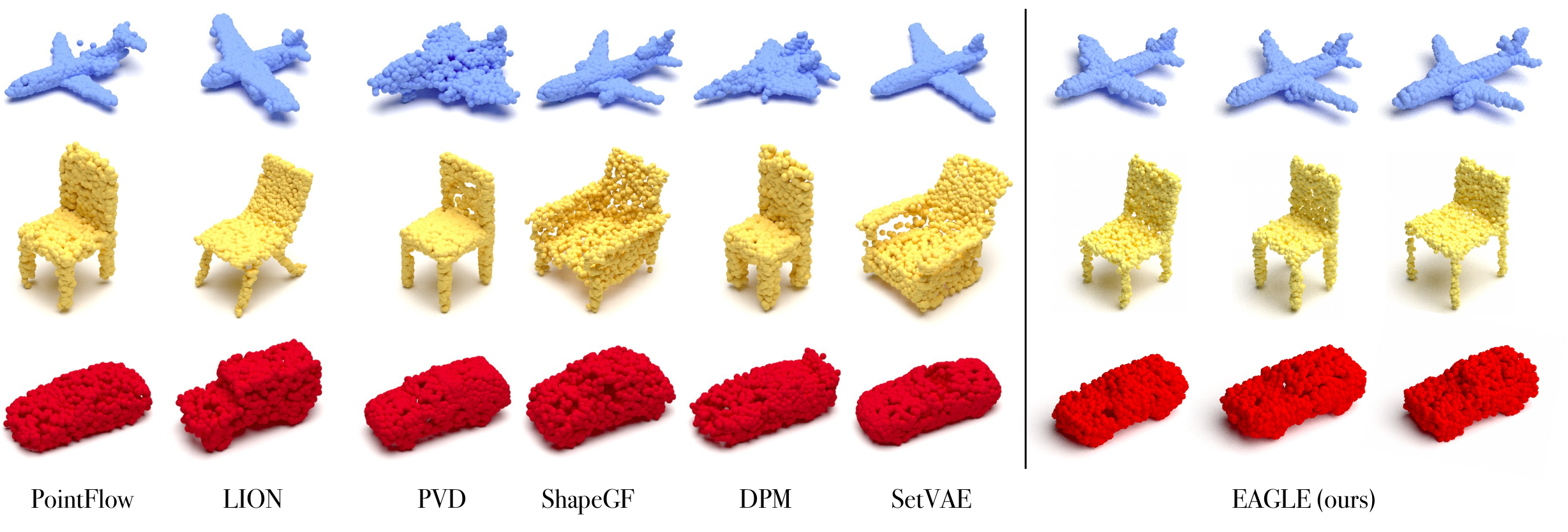}
    \caption{Visualization of our generation results (EAGLE) compared to baselines. EAGLE generates comparable and high-quality point clouds.}
    \label{fig:gen}
\end{figure*}
\begin{equation}
\frac{d\omega(t)}{dt} = f_\eta(\omega(t), t),
\label{eq:prior_cnf}
\end{equation}
where \(f_\eta(\cdot)\) denotes the unconditional vector field parameterized by \(\eta\),
\begin{equation}
\frac{dy(t)}{dt} = g_\alpha(y(t), c, t),
\label{eq:decoder_cnf}
\end{equation}
where \(g_\alpha(\cdot)\) represents the conditional vector field parameterized by \(\alpha\).
\paragraph{\textbf{Analysis of Global Drift in Conditional CNFs.}}
We observe that in conditional CNFs, the vector field integrates both temporal
and contextual information, which is typically achieved through context-dependent
bias terms. In the original implementation, the context-dependent bias is computed directly from the context vector denoted as 
\begin{align}
\mathbf{b}(c) = \phi (c),
\label{eq:std_bias}
\end{align}
where \(\phi(\cdot)\) denotes a learnable linear hyper-network (without bias term). 
For analysis, we consider a first-order approximation of the vector field expressed as
\begin{equation}
g_\theta(y, t, c) \approx W y(t) + \mathbf{b}(c),
\end{equation}
Mathematically, the solution to the ODE in Eq.~\ref{eq:decoder_cnf} over an integration interval \([0, T]\) can be approximated as
\begin{equation}
y(T) \approx y(0) + \int_0^T \left( W y(t) + \mathbf{b}(c) \right) dt.
\end{equation}
In practice, the context embedding is not guaranteed to be zero-centered, 
thus the expected bias generally satisfies \(\mathbb{E}[\mathbf{b}(c)] = \mu_b \neq 0\). Therefore, this constant component accumulates over time, resulting in a global shift \(T \cdot \mu_b\). 
This dominates the vector field in the early stages of ODE integration, causing a \emph{global drift issue}.
As this drift is continuously accumulated over time, the resulting flow
deviates from the target data manifold, leading to disappointing generation performance. We also empirically validate the global drift issue in conditional CNFs and the efficacy of our proposed mechanism in Sec. \ref{sec:EAGD}.
\paragraph{\textbf{Adaptive Bias Correction Mechanism.}}
To address this issue, we improve the fundamental neural ODE layer by introducing an
adaptive bias correction mechanism for contextual conditioning.
Specifically, we extend the base layer by augmenting the context-dependent bias
with layer normalization controlled by a learnable scaling factor.
Given a conditioning context \(c\), the corrected bias is defined as:
\begin{equation}
\mathbf{b}(c) = \gamma \cdot \phi(\mathrm{LN}(c)),
\label{eq:adaptive_bias}
\end{equation}
\(\mathrm{LN}(\cdot)\) is a layer normalization operation, and
\(\gamma\) is a learnable scalar that adaptively controls the strength of bias correction. 
Layer normalization is defined as:
\begin{equation}
\mathrm{LN}(c) = \frac{c - \mu}{\sqrt{\sigma^2 + \epsilon}},
\end{equation}
where \(\mu\) and \(\sigma^2\) denote the mean and variance of \(c\), and
\(\epsilon\) is a small constant for numerical stability.
Substituting this formulation into Eq.~\ref{eq:adaptive_bias}, the bias correction can be
written as:
\begin{equation}
\mathbf{b}(c)
= \gamma \cdot \phi \left( \frac{c - \mu}{\sqrt{\sigma^2 + \epsilon}} \right).
\label{eq:bias_final}
\end{equation}
The adaptive bias formulation is applied to the decoder CNF, where explicit conditioning on the latent variable is required.
For each ODE layer in the decoder CNF, the conditioned vector field is defined as
\begin{equation}
g_\alpha(y(t), z, t)
= \sigma\!\left( Wy + \mathbf{b}(c) \right),
\quad
c = [t, z],
\label{eq:g_alpha}
\end{equation}
where \(\sigma(\cdot)\) denotes a nonlinear activation function. The context \(c = [t, z]\) is formed by concatenating the time variable
\(t\) and the global shape latent \(z\) extracted by the point encoder, enabling joint temporal and latent
conditioning of the vector field.
\paragraph{\textbf{Effectiveness.}} The layer normalization operation eliminates the global mean of the context vector before it enters the hyper-network \(\phi\). 
Consequently, the expected value of the corrected bias approaches zero, i.e., \(\mathbb{E}[\mathbf{b}(c)] \approx 0\), 
which theoretically mitigates the early-stage drift caused by bias accumulation. 
Meanwhile, the adaptive scale \(\gamma\) allows the model to control the bias magnitude dynamically without reintroducing mean shifts. 

\subsection{Training Objective}
We adopt the same loss function as PointFlow \cite{yang2019pointflow}, which consists of three parts: reconstruction likelihood $L_{recon}$, prior $L_{prior}$, and posterior entropy $L_{entro}$.

$L_{recon}: E_{Q_\varphi(z | x)}[\log P_\alpha(X|z)]$ is the reconstruction log-likelihood of the input point data, which can be computed by Eq. \ref{eq:reconstruction}.

$L_{prior}: E_{Q_\varphi(z | x)}[\log P_\eta(z)]$ is used to make the encoded shape representation better adhere to the prior distribution, which is modeled by Eq. \ref{eq:prior}.

$L_{entro}: H\left[Q_\varphi(z | X)\right] $ measures the uncertainty or disorder of the latent variable $z$ under the approximated posterior distribution.

The final training objective $L_{total}$ can be summarized as 
\begin{align}
L_{total} = L_{recon} + L_{prior} + L_{entro}.
\end{align}
\begin{table*}[t]
\centering
\caption{Generation results on Airplane, Car and Chair compared with baselines using MMD, COV and JSD. MMD-CD and MMD-EMD scores are multiplied by $10^3$ and $10^2$, respectively; JSD is multiplied by $10^2$. Missing entries are denoted as “–”.
[Key: \textcolor{red}{best}, \textcolor{blue}{second best}]}
\label{tab:other_metric_results}
\scalebox{0.8}{
\begin{tabular}{l|ccccc|ccccc|ccccc}
\toprule
\multirow{3}{*}{Method}
& \multicolumn{5}{c}{Chair}
& \multicolumn{5}{c}{Airplane}
& \multicolumn{5}{c}{Car} \\
\cmidrule(lr){2-6} \cmidrule(lr){7-11} \cmidrule(lr){12-16}
& \multicolumn{2}{c}{MMD $\downarrow$}
& \multicolumn{2}{c}{COV(\%) $\uparrow$}
& JSD $\downarrow$
& \multicolumn{2}{c}{MMD $\downarrow$}
& \multicolumn{2}{c}{COV(\%) $\uparrow$}
& JSD $\downarrow$
& \multicolumn{2}{c}{MMD $\downarrow$}
& \multicolumn{2}{c}{COV(\%) $\uparrow$}
& JSD $\downarrow$ \\

& CD & EMD & CD & EMD & --
& CD & EMD & CD & EMD & --
& CD & EMD & CD & EMD & -- \\
\midrule

r-GAN \cite{achlioptas2018learning}
& 2.57 & 12.80 & 33.99 & 9.97 & 11.50
& 0.261 & 5.47 & 42.72 & 18.02 & 7.44
& 1.27 & 8.74 & 15.06 & 9.38 & 12.80 \\

l-GAN (CD) \cite{achlioptas2018learning}
& 2.46 & 8.91 & 41.39 & 25.68 & 4.59
& 0.239 & 4.27 & 43.21 & 21.23 & 4.62
& 1.55 & 6.25 & 38.64 & 18.47 & 4.43 \\

l-GAN (EMD) \cite{achlioptas2018learning}
& 2.61 & \textcolor{blue}{7.85} & 40.79 & 41.69 & 2.27
& 0.269 & 3.29 & 47.90 & 50.62 & \textcolor{red}{3.61}
& 1.48 & 5.43 & 39.20 & 39.77 & 2.21 \\

DPF-Net \cite{klokov2020discrete}
& 2.54 & -- & 44.71 & 48.79 & --
& 0.264 & -- & 46.17 & 48.89 & --
& 1.13 & -- & \textcolor{blue}{45.74} & 49.43 & -- \\

SoftFlow \cite{kim2020softflow}
& 2.53 & -- & 41.39 & 47.43 & --
& 0.231 & -- & 46.91 & 47.90 & --
& 1.19 & -- & 42.90 & 44.60 & -- \\

PointFlow \cite{yang2019pointflow}
& \textcolor{blue}{2.42} & 7.87 & 46.83 & 46.98 & \textcolor{blue}{1.74}
& \textcolor{blue}{0.217} & \textcolor{blue}{3.24} & 46.91 & 48.40 & 4.92
& \textcolor{blue}{0.91} & \textcolor{blue}{5.22} & 44.03 & 46.59 & \textcolor{blue}{0.87} \\

PC-GAN \cite{li2018point}
& 2.75 & 8.20 & 36.50 & 38.98 & 3.90
& 0.287 & 3.57 & 36.46 & 40.94 & 4.63
& 1.12 & 5.83 & 23.56 & 30.29 & 5.85 \\

PVD \cite{zhou20213d}
& 2.62 & -- & \textcolor{blue}{48.84} & \textcolor{blue}{50.60} & --
& 0.224 & -- & \textcolor{blue}{48.88} & \textcolor{blue}{52.09} & --
& 1.10 & -- & 41.19 & \textcolor{blue}{50.56} & -- \\
GET3D \cite{gao2022get3d}
& -- & -- & 43.36 & 42.77 & --
& -- & -- & -- & -- & --
& -- & -- & 15.04 & 18.38 & -- \\
MeshDiffusion \cite{Liu2023MeshDiffusion}
& -- & -- & 46.00 & 46.71 & --
& -- & -- & 47.34 & 42.15 & --
& -- & -- & 34.07 & 25.85 & -- \\
\midrule
EAGLE (ours)
& \textcolor{red}{2.35} & \textcolor{red}{7.73} & \textcolor{red}{51.19} & \textcolor{red}{51.30} & \textcolor{red}{1.53}
& \textcolor{red}{0.216} & \textcolor{red}{3.11} & \textcolor{red}{52.84} & \textcolor{red}{52.59} & \textcolor{blue}{4.61}
& \textcolor{red}{0.90} & \textcolor{red}{5.15} & \textcolor{red}{49.72} & \textcolor{red}{52.66} & \textcolor{red}{0.79} \\

\bottomrule
\end{tabular}
}
\end{table*}
\section{Experiments}
In this section, we first describe the experimental setup, followed by a controlled experiment to empirically validate the global drift issue and the efficacy of our proposed mechanism. We then compare our method with previous state-of-the-art models on point cloud generation task. Moreover, we evaluate the representation learning ability of the auto-encoder of our model. Finally, ablation studies demonstrate the rationality of the module designs.
\subsection{Experimental Setup}
\label{sec:Evaluation Metrics}
\noindent \textbf{Evaluation Metrics.} For comprehensive evaluation of generation quality, we employ the Chamfer Distance (CD) and Earth Mover’s Distance (EMD) as distance metrics to calculate
Coverage (COV), Minimum Matching Distance (MMD) and 1-Nearest Neighbor Accuracy (1-NNA). Jensen-Shannon Divergence (JSD) is also employed following \cite{yang2019pointflow}.
\begin{itemize}
\item \textbf{COV} measures how many real point clouds are covered by the generated point clouds, thus detecting mode collapse. However, While it fails to evaluate the quality of generated point clouds.
\item \textbf{MMD} used as a complementary metric to COV, mainly evaluates the matching quality between generated and real point cloud sets, measuring the fidelity of the generated point clouds \cite{luo2021diffusion}. But MMD is actually very insensitive to low-quality point clouds in the generated sets.
\item \textbf{JSD} compares the marginal distribution of generated point clouds with real point clouds by merging all generated or real ones and mapping them onto a voxel grid to discretize point distributions. However, JSD can not capture shape-level geometric coherence.
\item \textbf{1-NNA} measures the similarity between generated and real point clouds based on the error rate of a nearest-neighbor classifier by a 1-NN classifier,  which is more robust than the metrics mentioned above and correlates strongly with generation diversity and quality. For edge cases, when the generated point clouds and reference samples are identical, the 1-NNA classifier error rate will approach 50\%, indicating a good approximation of the target distribution \cite{klokov2020discrete}.
\end{itemize}
\begin{table}
    \centering
     \caption{Generation results on Airplane, Car, and Chair compared with baselines using 1-NNA  as the metric. Lower is better. For clarity, the table is divided into two splits: the bottom corresponds to  results from the past three years while the top corresponds to the earlier methods. [Key: \textcolor{red}{best}, \textcolor{blue}{second best}]}
    \label{tab:1-NN_generation_results}
    \scalebox{0.75}{
    \begin{tabular}{lcccccc} 
        \toprule
        \multirow{2}{*}{Method} & \multicolumn{2}{c}{Airplane} & \multicolumn{2}{c}{Car} & \multicolumn{2}{c}{Chair} \\
        \cmidrule(lr){2-3} \cmidrule(lr){4-5} \cmidrule(lr){6-7} & CD & EMD & CD & EMD & CD & EMD \\
        \midrule
        r-GAN\cite{achlioptas2018learning}  & 98.40 & 96.79 & 94.46 & 99.01 & 83.69 & 99.70 \\
        l-GAN (CD)\cite{achlioptas2018learning}  & 87.30 & 93.95 & 66.49 & 88.78 & 68.58 & 83.84 \\
        l-GAN (EMD)\cite{achlioptas2018learning}  & 89.49 & 76.91 & 71.16 & 66.19 & 71.90 & 64.65 \\
        PC-GAN \cite{li2018point} & 94.35 & 92.32 & 92.19 & 90.87 & 76.03 & 78.37 \\
        PointFlow \cite{yang2019pointflow} &  75.68 & 70.74 & 60.65 & 62.36 & 62.84 & 60.57 \\
        SoftFlow \cite{kim2020softflow} & 76.05 & 65.80 & 62.35 & 54.48 & 59.21 & 60.05 \\
        DPF-Net \cite{klokov2020discrete}& 75.18 & 65.55 & 62.35 & 54.48 & 62.00 & 58.53 \\
        SetVAE \cite{kim2021setvae} & 76.54 & 67.65 & 59.95 & 59.94 & 58.84 & 60.57 \\
        DPM \cite{luo2021diffusion}& 76.42 & 86.91 & 68.89 & 79.97 & 60.05 & 74.77 \\
        Shape-GF \cite{cai2020learning} & 80.00 & 76.17 & 63.20 & 56.53 & 68.96 & 65.48 \\
        CanonicalVAE \cite{cheng2022autoregressive} & 80.15 & 76.27 & 63.23 & 61.56 & 62.78 & 61.05 \\
        PVD \cite{zhou20213d}& \textcolor{blue}{73.82} & \textcolor{red}{64.81} & \textcolor{red}{54.55} & \textcolor{blue}{53.83} & \textcolor{red}{56.26} & \textcolor{red}{53.32} \\
        GET3D \cite{gao2022get3d}& -- & -- & 75.26 & 72.49 & 75.26 & 72.49 \\
        SPAGHETTI \cite{hertz2022spaghetti}& 78.20 & 77.00 & 72.30 & 71.00 & 70.70 & 69.00 \\
        EAGLE (ours) & \textcolor{red}{71.85} & \textcolor{blue}{65.02} & \textcolor{blue}{57.39} & \textcolor{red}{53.26} & \textcolor{blue}{58.39} & \textcolor{blue}{58.16} \\
        \midrule
        PointGPT \cite{chen2023pointgpt} & 74.85 & 65.61 & \textcolor{red}{55.91} & 54.24 & 57.24 & \textcolor{red}{55.01} \\
        MeshDiffusion \cite{Liu2023MeshDiffusion}  & \textcolor{red}{66.44} & 76.26 & 81.43 & 87.84 & \textcolor{red}{53.69} & \textcolor{blue}{57.63} \\
        SALAD \cite{koo2023salad}  & 73.90 & 71.10 & 59.20 & 57.20 & 57.80 & 58.40 \\
        % PCGen \cite{vercheval2024pcgen}& 69.01 & 68.64 & 57.81 & 56.67 & 58.45 & 58.91 \\
        % Diff-PCG \cite{yu2025diff} & \textcolor{red}{65.12} & 75.45 & 66.50 & 69.27 & 55.62 & 62.31 \\
        NSOT \cite{hui2025notsooptimal}  & \textcolor{blue}{68.64} & \textcolor{red}{61.85} & 59.66 & \textcolor{blue}{53.55} & \textcolor{blue}{55.51} & \textcolor{blue}{57.63} \\
        DiPT \cite{bastico2026rethinking} & -- & 74.32 & -- & 60.26 & -- & 64.47 \\
        EAGLE (ours) & 71.85 & \textcolor{blue}{65.02} & \textcolor{blue}{57.39} & \textcolor{red}{53.26} & 58.39 & 58.16 \\
        \bottomrule
    \end{tabular}
    }
\end{table}

\noindent \textbf{Configuration.} For the generation task, we follow previous works \cite{yang2019pointflow,cai2020learning,kim2020softflow,klokov2020discrete,zhou20213d} and select three classes from ShapeNet dataset \cite{chang2015shapenet}: airplane, chair, and car. Each class in ShapeNet contains 15,000 sampled points. We sample 2,048 points for training and testing for each shape and preprocess the point cloud data following the steps outlined in \cite{yang2019pointflow}. For the classification evaluation of our auto-encoders, we implement ModelNet40 and ModelNet10 datasets \cite{wu20153d}, which contain 3D models of 40 and 10 categories, respectively.
\begin{figure}
    \centering
\includegraphics[width=1\linewidth]{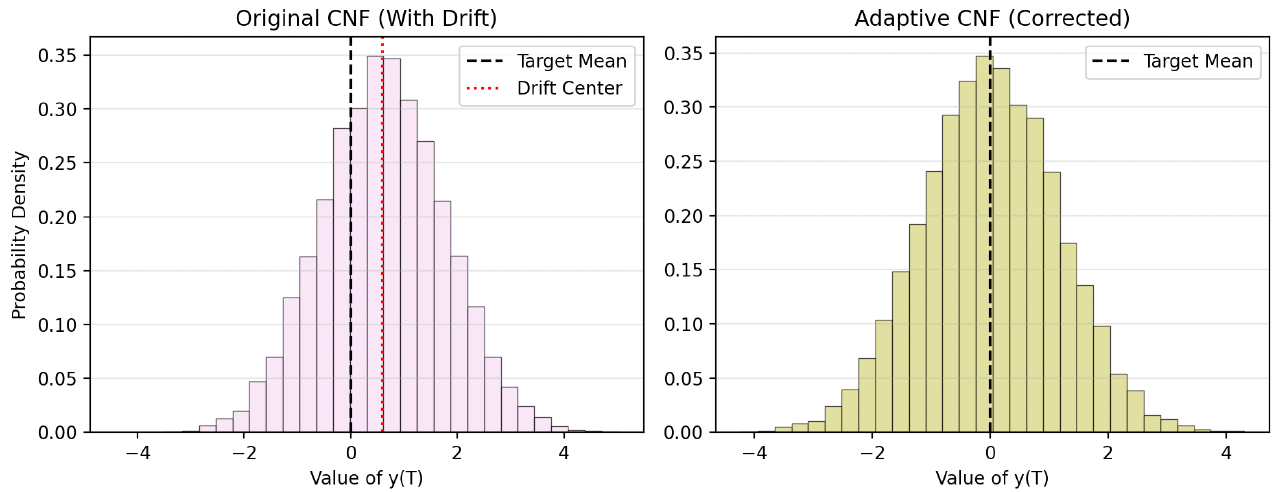}
    \caption{Empirical validation of the global drift issue and our correction.
    Both histograms show the distribution of the final ODE state $y(T)$ after integration over $t \in [0, 1]$, starting from $y(0) \sim \mathcal{N}(0, 1)$. The x-axis denotes the mean value of the generated sample $y(T)$, where the target mean is 0 (black dashed line).
    The y-axis represents probability density.}
    \label{fig:proof}
\end{figure}
\subsection{Empirical Analysis of Global Drift}
\label{sec:EAGD}
To empirically validate the global drift issue in conditional CNFs and the efficacy of our proposed mechanism, we conduct a controlled experiment shown in Fig. \ref{fig:proof} using real context embeddings extracted from the trained encoder. Specifically, we extract the global shape latent $z$, which is then concatenated with the time variable $t$ to form the conditioning context $c = [t, z]$. This $c$ is identical to the context used during standard training and inference of the decoder CNF. Then, we draw $N = 10{,}000$ independent samples from the initial distribution $y(0) \sim \mathcal{N}(0, I)$ and condition them on context vectors $c$. This large sample size ensures a stable and low-variance empirical estimation of the final state distribution $p(y(T))$, enabling reliable quantification and clear visualization of distributional differences between the original and adaptive CNFs.
\begin{figure*}[t]
    \centering
\includegraphics[width=\textwidth]{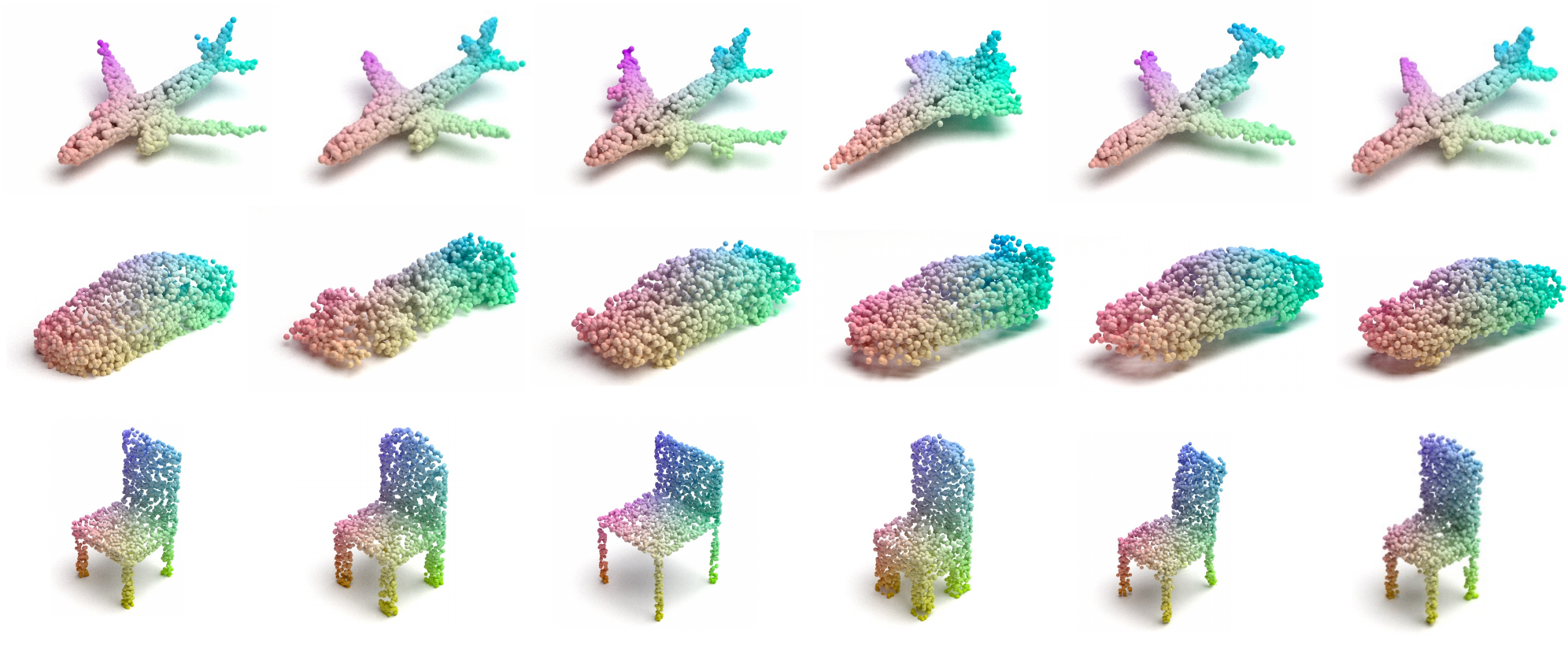}
    \caption{More visualization results of point clouds generated by our model. From top to bottom: airplane, car, chair.}
    \label{fig:gen_all}
\end{figure*}
Starting from noise samples $y(0) \sim \mathcal{N}(0, I)$, we integrate the ODEs over $t \in [0, 1]$ to obtain the final generated states $y(T)$. As shown in Fig.~\ref{fig:proof} (left), the original CNF exhibits a pronounced \emph{global mean drift}. The empirical mean of its output distribution $y(T)$ deviates significantly from the target origin. As this drift continuously accumulates over time, this offset shifts the entire generated distribution away from the desired data manifold, leading to disappointing generation performance. Fig.~\ref{fig:proof} (right) demonstrates that adaptive CNFs effectively stabilize the ODE integration process compared to the original bias formulation.
\subsection{Point Cloud Generation}
\paragraph{\textbf{Qualitative comparisons}} The qualitative results shown in Fig. \ref{fig:gen} demonstrate that our generation results have achieved competitive quality compared to other baseline methods. The visualization of the baseline methods is taken from LION \cite{vahdat2022lion}. Fig. \ref{fig:gen_all} shows more visualization results generated by our model. 
\paragraph{\textbf{Quantitative comparisons}} We quantitatively compare our generated results with previous state-of-art methods using the evaluation metrics described in Sec. \ref{sec:Evaluation Metrics}. We summarize the results in Tab. \ref{tab:other_metric_results} and Tab. \ref{tab:1-NN_generation_results}. 
As shown in Tab.~\ref{tab:other_metric_results}, our method achieves almost the best performance across all three categories. Tab.~\ref{tab:1-NN_generation_results} further shows that EAGLE obtains competitive 1-NNA scores, demonstrating strong distribution similarity between generated point clouds and ground truth.

We note that fewer comparisons are reported in Tab.~\ref{tab:other_metric_results}, as several methods do not provide official results for MMD, COV or JSD. In contrast, 1-NNA has recently become a more widely adopted metric as shown in Tab. \ref{tab:1-NN_generation_results}.

\begin{table}[t]
\centering
\caption{Classification results on ModelNet10 (MN-10) and ModelNet40 (MN-40) datasets. Missing entries are denoted as “–”.}
\label{tab:modelnet_cls}
\scalebox{1}{
\begin{tabular}{lcc}
\toprule
Model & MN-10 (\%) & MN-40 (\%) \\
\midrule
SPH\cite{kazhdan2003rotation}          & 68.2 & 79.8 \\
LFD\cite{chen2003visual}          & 75.5 & 79.9 \\
T-L Network\cite{girdhar2016learning}  & 74.4 & --   \\
Vconv-DAE\cite{sharma2016vconv}    & 75.5 & 80.5 \\
PointGrow\cite{sun2020pointgrow}    & 85.7 & --   \\
MRTNet-VAE\cite{gadelha2018multiresolution}   & 86.4 & --   \\
3D-GAN\cite{wu2016learning}       & 83.3 & 91.0 \\
PointFlow\cite{yang2019pointflow}    & 86.8 & 93.7 \\
DPM\cite{luo2021diffusion}          & 87.6 & \textbf{94.2} \\
DualGAN\cite{wen2021learning}       &87.6  & 94.1 \\
TGN\cite{xu2023transformer}          & \textbf{87.7} & \textbf{94.2} \\
\midrule
EAGLE (ours) & \textbf{87.7} & 93.5 \\
\bottomrule
\end{tabular}
}
\end{table}
\subsection{Representation Learning Evaluation}
To additionally evaluate the representation learning ability of auto-encoders, we conduct classification experiments following previous methods \cite{yang2019pointflow,xu2023transformer}. We first train our proposed model with the full ShapeNet dataset and then train a linear SVM with extracted latent representations for classification. The classification results are summarized in Tab \ref{tab:modelnet_cls}. Our method achieves the same performance as TGN \cite{xu2023transformer} and outperforms other methods on the ModelNet10 dataset, while demonstrating comparable results on the ModelNet40 dataset.

\subsection{Ablation Study}
During ablation studies, we consider PointFlow \cite{yang2019pointflow} as the vanilla and take the mean of 1-NNA (CD / EMD) scores across three categories (airplane, car and chair), respectively. We evaluate the impact of the foundational layers (ConcatLinear or Adaptive-ConcatLinear), residual block and self-attention mechanism on model performance. Starting from a basic vanilla, we progressively incorporate each proposed module, and observe consistent performance improvements almost at each stage. Notably, no performance degradation is observed when introducing additional modules. The experimental results from different combinations of components presented in Tab. \ref{tab:ablation} validate the individual contribution of each component as well as their complementary effects when combined.
\begin{table}
    \centering
    \caption{Ablation of foundational layers, residual block, and self-attention using 1-NNA as the metric. [SA = Self-Attention; RB = Residual Block; CL = ConcatLinear of CNFs adopted by PointFlow \cite{yang2019pointflow}; ACL = Adaptive-ConcatLinear of A-CNFs adopted by EAGLE; Key: \textcolor{red}{best}, \textcolor{blue}{second best}]}
    \label{tab:ablation}
    \scalebox{0.8}{
    \begin{tabular}{c|c|c|c|c|cc}
        \toprule
        Method & ACL & CL &  RB & SA & Mean CD $\downarrow$ & Mean EMD $\downarrow$ \\
        \midrule
        vanilla & -- & \checkmark &   -- & -- & 66.39 & 64.56 \\
        \midrule
        Ours & \checkmark & -- &   -- & -- & 65.79 & 64.02 \\
        \midrule
        Ours & -- & \checkmark &   -- & \checkmark & 63.85 & 61.88
        \\
        \midrule
        Ours & \checkmark & -- &   -- & \checkmark & {63.11} & \textcolor{blue}{59.87} \\
        \midrule
        Ours & -- & \checkmark &   \checkmark & \checkmark & \textcolor{blue}{63.01} & {59.97} \\
        \midrule
        Ours(EAGLE) & \checkmark & -- &   \checkmark & \checkmark & \textcolor{red}{62.74} & \textcolor{red}{58.81} \\
        \bottomrule
    \end{tabular}
    }
\end{table}
\section{Conclusion}
In this paper, we introduce a novel probabilistic generative model that incorporates self-attention mechanism to effectively model long-range dependencies and global structural relationships
among points, enabling more coherent and semantically meaningful generation. We also propose adaptive continuous normalizing flows by introducing an adaptive bias correction mechanism accompanied by a detailed empirical analysis to better provide deeper insights into the global drift issue. Extensive experimental results demonstrate that the proposed method achieves competitive or even best performance across multiple quantitative metrics. Ablation studies further provide strong evidence for the effectiveness and necessity of each proposed module.\\\\
\textbf{Limitations and Future Work}
Although our proposed method is capable of generating high-quality point clouds, it comes with significant computational challenges. The primary issue arises from the extensive solving of ordinary differential equations (ODEs) \cite{chen2018neural} during the training and inference phases. In contrast, discrete flows, such as RealNVP \cite{dinh2016density} and Glow \cite{kingma2018glow}, are invertible mappings that have computational efficiency, making them much faster compared to continuous flows. However, their ability to model complex, high-dimensional distributions is somewhat limited due to the relatively simple nature of the transformations they employ. Therefore, we consider designing a hybrid flow framework that integrates the advantages of both discrete flows and continuous normalizing flows, where the discrete flow performs fast, large-scale transformations for the initial stages of the transformation process, then the continuous normalizing flow is used for detailed probability density transformations. Adopting the aforementioned approach for point cloud generation and related tasks may effectively alleviate the issue of computational complexity.

\section*{Acknowledgements}
This work was supported by the National Natural Science Foundation of China under Grant 62302280 and 62472265.

%%%%%%%%% REFERENCES
{\small
\bibliographystyle{ieee_fullname}
\bibliography{egbib}

@String(ECCV= {Eur. Conf. Comput. Vis.})

@String(TOG= {ACM Trans. Graph.})

@String(ECCV  = {ECCV})

@String(TOG   = {ACM TOG})

@inproceedings{achlioptas2018learning,
  title={Learning representations and generative models for 3d point clouds},
  author={Achlioptas, Panos and Diamanti, Olga and Mitliagkas, Ioannis and Guibas, Leonidas},
  booktitle={International conference on machine learning},
  pages={40--49},
  year={2018},
  organization={PMLR}
}

@inproceedings{gadelha2018multiresolution,
  title={Multiresolution tree networks for 3d point cloud processing},
  author={Gadelha, Matheus and Wang, Rui and Maji, Subhransu},
  booktitle={Proceedings of the European Conference on Computer Vision (ECCV)},
  pages={103--118},
  year={2018}
}

@article{wu2016learning,
  title={Learning a probabilistic latent space of object shapes via 3d generative-adversarial modeling},
  author={Wu, Jiajun and Zhang, Chengkai and Xue, Tianfan and Freeman, Bill and Tenenbaum, Josh},
  journal={Advances in neural information processing systems},
  volume={29},
  year={2016}
}

@article{gulrajani2017improved,
  title={Improved training of wasserstein gans},
  author={Gulrajani, Ishaan and Ahmed, Faruk and Arjovsky, Martin and Dumoulin, Vincent and Courville, Aaron C},
  journal={Advances in neural information processing systems},
  volume={30},
  year={2017}
}

@article{mo2023dit,
  title={Dit-3d: Exploring plain diffusion transformers for 3d shape generation},
  author={Mo, Shentong and Xie, Enze and Chu, Ruihang and Hong, Lanqing and Niessner, Matthias and Li, Zhenguo},
  journal={Advances in neural information processing systems},
  volume={36},
  pages={67960--67971},
  year={2023}
}

@article{goodfellow2014generative,
  title={Generative adversarial nets},
  author={Goodfellow, Ian and Pouget-Abadie, Jean and Mirza, Mehdi and Xu, Bing and Warde-Farley, David and Ozair, Sherjil and Courville, Aaron and Bengio, Yoshua},
  journal={Advances in neural information processing systems},
  volume={27},
  year={2014}
}

@inproceedings{qi2017pointnet,
  title={Pointnet: Deep learning on point sets for 3d classification and segmentation},
  author={Qi, Charles R and Su, Hao and Mo, Kaichun and Guibas, Leonidas J},
  booktitle={Proceedings of the IEEE conference on computer vision and pattern recognition},
  pages={652--660},
  year={2017}
}

@inproceedings{zhou20213d,
  title={3d shape generation and completion through point-voxel diffusion},
  author={Zhou, Linqi and Du, Yilun and Wu, Jiajun},
  booktitle={Proceedings of the IEEE/CVF international conference on computer vision},
  pages={5826--5835},
  year={2021}
}

@inproceedings{arjovsky2017wasserstein,
  title={Wasserstein generative adversarial networks},
  author={Arjovsky, Martin and Chintala, Soumith and Bottou, L{\'e}on},
  booktitle={International conference on machine learning},
  pages={214--223},
  year={2017},
  organization={PMLR}
}

@inproceedings{sharma2016vconv,
  title={Vconv-dae: Deep volumetric shape learning without object labels},
  author={Sharma, Abhishek and Grau, Oliver and Fritz, Mario},
  booktitle={European conference on computer vision},
  pages={236--250},
  year={2016},
  organization={Springer}
}

@inproceedings{girdhar2016learning,
  title={Learning a predictable and generative vector representation for objects},
  author={Girdhar, Rohit and Fouhey, David F and Rodriguez, Mikel and Gupta, Abhinav},
  booktitle={European conference on computer vision},
  pages={484--499},
  year={2016},
  organization={Springer}
}

@inproceedings{wen2021learning,
  title={Learning progressive point embeddings for 3d point cloud generation},
  author={Wen, Cheng and Yu, Baosheng and Tao, Dacheng},
  booktitle={Proceedings of the IEEE/CVF Conference on Computer Vision and Pattern Recognition},
  pages={10266--10275},
  year={2021}
}

@article{ho2020denoising,
  title={Denoising diffusion probabilistic models},
  author={Ho, Jonathan and Jain, Ajay and Abbeel, Pieter},
  journal={Advances in neural information processing systems},
  volume={33},
  pages={6840--6851},
  year={2020}
}

@inproceedings{cheng2022autoregressive,
  title={Autoregressive 3d shape generation via canonical mapping},
  author={Cheng, An-Chieh and Li, Xueting and Liu, Sifei and Sun, Min and Yang, Ming-Hsuan},
  booktitle={European Conference on Computer Vision},
  pages={89--104},
  year={2022},
  organization={Springer}
}

@article{chen2023pointgpt,
  title={Pointgpt: Auto-regressively generative pre-training from point clouds},
  author={Chen, Guangyan and Wang, Meiling and Yang, Yi and Yu, Kai and Yuan, Li and Yue, Yufeng},
  journal={Advances in Neural Information Processing Systems},
  volume={36},
  pages={29667--29679},
  year={2023}
}

@inproceedings{xu2023transformer,
  title={Transformer-based point cloud generation network},
  author={Xu, Rui and Hui, Le and Han, Yuehui and Qian, Jianjun and Xie, Jin},
  booktitle={Proceedings of the 31st ACM International Conference on Multimedia},
  pages={4169--4177},
  year={2023}
}

@inproceedings{chen2003visual,
  title={On visual similarity based 3D model retrieval},
  author={Chen, Ding-Yun and Tian, Xiao-Pei and Shen, Yu-Te and Ouhyoung, Ming},
  booktitle={Computer graphics forum},
  volume={22},
  pages={223--232},
  year={2003},
  organization={Wiley Online Library}
}

@inproceedings{yang2019pointflow,
  title={Pointflow: 3d point cloud generation with continuous normalizing flows},
  author={Yang, Guandao and Huang, Xun and Hao, Zekun and Liu, Ming-Yu and Belongie, Serge and Hariharan, Bharath},
  booktitle={Proceedings of the IEEE/CVF international conference on computer vision},
  pages={4541--4550},
  year={2019}
}

@inproceedings{luo2021diffusion,
  title={Diffusion probabilistic models for 3d point cloud generation},
  author={Luo, Shitong and Hu, Wei},
  booktitle={Proceedings of the IEEE/CVF conference on computer vision and pattern recognition},
  pages={2837--2845},
  year={2021}
}

@article{gao2022get3d,
  title={Get3d: A generative model of high quality 3d textured shapes learned from images},
  author={Gao, Jun and Shen, Tianchang and Wang, Zian and Chen, Wenzheng and Yin, Kangxue and Li, Daiqing and Litany, Or and Gojcic, Zan and Fidler, Sanja},
  journal={Advances in neural information processing systems},
  volume={35},
  pages={31841--31854},
  year={2022}
}

@inproceedings{groueix2018papier,
  title={A papier-m{\^a}ch{\'e} approach to learning 3d surface generation},
  author={Groueix, Thibault and Fisher, Matthew and Kim, Vladimir G and Russell, Bryan C and Aubry, Mathieu},
  booktitle={Proceedings of the IEEE conference on computer vision and pattern recognition},
  pages={216--224},
  year={2018}
}

@inproceedings{kazhdan2003rotation,
  title={Rotation invariant spherical harmonic representation of 3 d shape descriptors},
  author={Kazhdan, Michael and Funkhouser, Thomas and Rusinkiewicz, Szymon},
  booktitle={Symposium on geometry processing},
  volume={6},
  pages={156--164},
  year={2003}
}

@article{zhang2025enhancing,
  title={Enhancing 3D point cloud generation via Mamba-based time-varying denoising diffusion},
  author={Zhang, Daopeng and Yu, Li},
  journal={Journal of Visual Communication and Image Representation},
  pages={104657},
  year={2025},
  publisher={Elsevier}
}

@inproceedings{kim2021setvae,
  title={Setvae: Learning hierarchical composition for generative modeling of set-structured data},
  author={Kim, Jinwoo and Yoo, Jaehoon and Lee, Juho and Hong, Seunghoon},
  booktitle={Proceedings of the IEEE/CVF Conference on Computer Vision and Pattern Recognition},
  pages={15059--15068},
  year={2021}
}

@article{zamorski2018adversarial,
  title={Adversarial autoencoders for generating 3d point clouds},
  author={Zamorski, Maciej and Zieba, Maciej and Nowak, Rafa{\l} and Stokowiec, Wojciech and Trzcinski, Tomasz},
  journal={arXiv preprint arXiv:1811.07605},
  volume={2},
  number={3},
  year={2018}
}

@inproceedings{koo2023salad,
  title={Salad: Part-level latent diffusion for 3d shape generation and manipulation},
  author={Koo, Juil and Yoo, Seungwoo and Nguyen, Minh Hieu and Sung, Minhyuk},
  booktitle={Proceedings of the ieee/cvf international conference on computer vision},
  pages={14441--14451},
  year={2023}
}

@article{hertz2022spaghetti,
  title={Spaghetti: Editing implicit shapes through part aware generation},
  author={Hertz, Amir and Perel, Or and Giryes, Raja and Sorkine-Hornung, Olga and Cohen-Or, Daniel},
  journal={ACM Transactions on Graphics (TOG)},
  volume={41},
  number={4},
  pages={1--20},
  year={2022},
  publisher={ACM New York, NY, USA}
}

@inproceedings{bastico2026rethinking,
  title={Rethinking Metrics and Diffusion Architecture for 3D Point Cloud Generation},
  author={Bastico, Matteo and Ryckelynck, David and Cort{\'e}, Laurent and Tillier, Yannick and Decenci{\`e}re, Etienne},
  booktitle={Thirteenth International Conference on 3D Vision},
  year={2026}
}

@inproceedings{
hui2025notsooptimal,
title={Not-So-Optimal Transport Flows for 3D Point Cloud Generation},
author={Ka-Hei Hui and Chao Liu and Xiaohui Zeng and Chi-Wing Fu and Arash Vahdat},
booktitle={The Thirteenth International Conference on Learning Representations},
year={2025},
url={https://openreview.net/forum?id=62Ff8LDAJZ}
}

@InProceedings{Liu2023MeshDiffusion,
    title={MeshDiffusion: Score-based Generative 3D Mesh Modeling},
    author={Zhen Liu and Yao Feng and Michael J. Black and Derek Nowrouzezahrai and Liam Paull and Weiyang Liu},
    booktitle={International Conference on Learning Representations},
    year={2023},
    url={https://openreview.net/forum?id=0cpM2ApF9p6}
}

@article{makhzani2015adversarial,
  title={Adversarial autoencoders},
  author={Makhzani, Alireza and Shlens, Jonathon and Jaitly, Navdeep and Goodfellow, Ian and Frey, Brendan},
  journal={arXiv preprint arXiv:1511.05644},
  year={2015}
}

@inproceedings{wu20153d,
  title={3d shapenets: A deep representation for volumetric shapes},
  author={Wu, Zhirong and Song, Shuran and Khosla, Aditya and Yu, Fisher and Zhang, Linguang and Tang, Xiaoou and Xiao, Jianxiong},
  booktitle={Proceedings of the IEEE conference on computer vision and pattern recognition},
  pages={1912--1920},
  year={2015}
}

@inproceedings{he2016deep,
  title={Deep residual learning for image recognition},
  author={He, Kaiming and Zhang, Xiangyu and Ren, Shaoqing and Sun, Jian},
  booktitle={Proceedings of the IEEE conference on computer vision and pattern recognition},
  pages={770--778},
  year={2016}
}

@article{li2018point,
  title={Point cloud gan},
  author={Li, Chun-Liang and Zaheer, Manzil and Zhang, Yang and Poczos, Barnabas and Salakhutdinov, Ruslan},
  journal={arXiv preprint arXiv:1810.05795},
  year={2018}
}

@article{grathwohl2018ffjord,
  title={Ffjord: Free-form continuous dynamics for scalable reversible generative models},
  author={Grathwohl, Will and Chen, Ricky TQ and Bettencourt, Jesse and Sutskever, Ilya and Duvenaud, David},
  journal={arXiv preprint arXiv:1810.01367},
  year={2018}
}

@article{dinh2014nice,
  title={Nice: Non-linear independent components estimation},
  author={Dinh, Laurent and Krueger, David and Bengio, Yoshua},
  journal={arXiv preprint arXiv:1410.8516},
  year={2014}
}

@article{vahdat2022lion,
  title={Lion: Latent point diffusion models for 3d shape generation},
  author={Vahdat, Arash and Williams, Francis and Gojcic, Zan and Litany, Or and Fidler, Sanja and Kreis, Karsten and others},
  journal={Advances in Neural Information Processing Systems},
  volume={35},
  pages={10021--10039},
  year={2022}
}

@article{chen2018neural,
  title={Neural ordinary differential equations},
  author={Chen, Ricky TQ and Rubanova, Yulia and Bettencourt, Jesse and Duvenaud, David K},
  journal={Advances in neural information processing systems},
  volume={31},
  year={2018}
}

@article{kingma2018glow,
  title={Glow: Generative flow with invertible 1x1 convolutions},
  author={Kingma, Durk P and Dhariwal, Prafulla},
  journal={Advances in neural information processing systems},
  volume={31},
  year={2018}
}

@inproceedings{klokov2020discrete,
  title={Discrete point flow networks for efficient point cloud generation},
  author={Klokov, Roman and Boyer, Edmond and Verbeek, Jakob},
  booktitle={European Conference on Computer Vision},
  pages={694--710},
  year={2020},
  organization={Springer}
}

@article{papamakarios2017masked,
  title={Masked autoregressive flow for density estimation},
  author={Papamakarios, George and Pavlakou, Theo and Murray, Iain},
  journal={Advances in neural information processing systems},
  volume={30},
  year={2017}
}

@article{chang2015shapenet,
  title={Shapenet: An information-rich 3d model repository},
  author={Chang, Angel X and Funkhouser, Thomas and Guibas, Leonidas and Hanrahan, Pat and Huang, Qixing and Li, Zimo and Savarese, Silvio and Savva, Manolis and Song, Shuran and Su, Hao and others},
  journal={arXiv preprint arXiv:1512.03012},
  year={2015}
}

@article{kingma2013auto,
  title={Auto-encoding variational bayes},
  author={Kingma, Diederik P},
  journal={arXiv preprint arXiv:1312.6114},
  year={2013}
}

@inproceedings{fan2017point,
  title={A point set generation network for 3d object reconstruction from a single image},
  author={Fan, Haoqiang and Su, Hao and Guibas, Leonidas J},
  booktitle={Proceedings of the IEEE conference on computer vision and pattern recognition},
  pages={605--613},
  year={2017}
}

@inproceedings{ren2024tiger,
  title={TIGER: Time-Varying Denoising Model for 3D Point Cloud Generation with Diffusion Process},
  author={Ren, Zhiyuan and Kim, Minchul and Liu, Feng and Liu, Xiaoming},
  booktitle={Proceedings of the IEEE/CVF Conference on Computer Vision and Pattern Recognition},
  pages={9462--9471},
  year={2024}
}

@article{guo2021pct,
  title={Pct: Point cloud transformer},
  author={Guo, Meng-Hao and Cai, Jun-Xiong and Liu, Zheng-Ning and Mu, Tai-Jiang and Martin, Ralph R and Hu, Shi-Min},
  journal={Computational Visual Media},
  volume={7},
  pages={187--199},
  year={2021},
  publisher={Springer}
}

@inproceedings{zhao2021point,
  title={Point transformer},
  author={Zhao, Hengshuang and Jiang, Li and Jia, Jiaya and Torr, Philip HS and Koltun, Vladlen},
  booktitle={Proceedings of the IEEE/CVF international conference on computer vision},
  pages={16259--16268},
  year={2021}
}

@article{kim2020softflow,
  title={Softflow: Probabilistic framework for normalizing flow on manifolds},
  author={Kim, Hyeongju and Lee, Hyeonseung and Kang, Woo Hyun and Lee, Joun Yeop and Kim, Nam Soo},
  journal={Advances in Neural Information Processing Systems},
  volume={33},
  pages={16388--16397},
  year={2020}
}

@article{wu2022point,
  title={Point transformer v2: Grouped vector attention and partition-based pooling},
  author={Wu, Xiaoyang and Lao, Yixing and Jiang, Li and Liu, Xihui and Zhao, Hengshuang},
  journal={Advances in Neural Information Processing Systems},
  volume={35},
  pages={33330--33342},
  year={2022}
}

@article{vaswani2017attention,
  title={Attention is all you need},
  author={Vaswani, Ashish and Shazeer, Noam and Parmar, Niki and Uszkoreit, Jakob and Jones, Llion and Gomez, Aidan N and Kaiser, {\L}ukasz and Polosukhin, Illia},
  journal={Advances in neural information processing systems},
  volume={30},
  year={2017}
}

@inproceedings{postels2021go,
  title={Go with the flows: Mixtures of normalizing flows for point cloud generation and reconstruction},
  author={Postels, Janis and Liu, Mengya and Spezialetti, Riccardo and Van Gool, Luc and Tombari, Federico},
  booktitle={2021 International Conference on 3D Vision (3DV)},
  pages={1249--1258},
  year={2021},
  organization={IEEE}
}

@article{papamakarios2021normalizing,
  title={Normalizing flows for probabilistic modeling and inference},
  author={Papamakarios, George and Nalisnick, Eric and Rezende, Danilo Jimenez and Mohamed, Shakir and Lakshminarayanan, Balaji},
  journal={Journal of Machine Learning Research},
  volume={22},
  number={57},
  pages={1--64},
  year={2021}
}

@inproceedings{rezende2015variational,
  title={Variational inference with normalizing flows},
  author={Rezende, Danilo and Mohamed, Shakir},
  booktitle={International conference on machine learning},
  pages={1530--1538},
  year={2015},
  organization={PMLR}
}

@inproceedings{yao2023local,
  title={Local implicit normalizing flow for arbitrary-scale image super-resolution},
  author={Yao, Jie-En and Tsao, Li-Yuan and Lo, Yi-Chen and Tseng, Roy and Chang, Chia-Che and Lee, Chun-Yi},
  booktitle={Proceedings of the IEEE/CVF Conference on Computer Vision and Pattern Recognition},
  pages={1776--1785},
  year={2023}
}

@inproceedings{yang2018foldingnet,
  title={Foldingnet: Point cloud auto-encoder via deep grid deformation},
  author={Yang, Yaoqing and Feng, Chen and Shen, Yiru and Tian, Dong},
  booktitle={Proceedings of the IEEE conference on computer vision and pattern recognition},
  pages={206--215},
  year={2018}
}

@article{dinh2016density,
  title={Density estimation using real nvp},
  author={Dinh, Laurent and Sohl-Dickstein, Jascha and Bengio, Samy},
  journal={arXiv preprint arXiv:1605.08803},
  year={2016}
}

@inproceedings{cai2020learning,
  title={Learning gradient fields for shape generation},
  author={Cai, Ruojin and Yang, Guandao and Averbuch-Elor, Hadar and Hao, Zekun and Belongie, Serge and Snavely, Noah and Hariharan, Bharath},
  booktitle={Computer Vision--ECCV 2020: 16th European Conference, Glasgow, UK, August 23--28, 2020, Proceedings, Part III 16},
  pages={364--381},
  year={2020},
  organization={Springer}
}

@inproceedings{sun2020pointgrow,
  title={Pointgrow: Autoregressively learned point cloud generation with self-attention},
  author={Sun, Yongbin and Wang, Yue and Liu, Ziwei and Siegel, Joshua and Sarma, Sanjay},
  booktitle={Proceedings of the IEEE/CVF Winter Conference on Applications of Computer Vision},
  pages={61--70},
  year={2020}
}

@inproceedings{wang2024neighborhood,
  title={Neighborhood Feature Enhancement Flow Diffusion Model for Point Cloud Generation},
  author={Wang, Hongcheng and Zhang, Dongdong and Liu, Taotao and Qi, Xumai},
  booktitle={International Conference on Pattern Recognition},
  pages={339--354},
  year={2024},
  organization={Springer}
}

@inproceedings{wu2023sketch,
  title={Sketch and text guided diffusion model for colored point cloud generation},
  author={Wu, Zijie and Wang, Yaonan and Feng, Mingtao and Xie, He and Mian, Ajmal},
  booktitle={Proceedings of the IEEE/CVF International Conference on Computer Vision},
  pages={8929--8939},
  year={2023}
}

@article{wang2024dpr,
  title={DPR-GAN: Dual-Stream Progressive Refinement for Adversarial 3D Point Cloud Generation},
  author={Wang, Xiangyang and Chen, Jiale and Wang, Rui},
  journal={Neural Processing Letters},
  volume={56},
  number={2},
  pages={70},
  year={2024},
  publisher={Springer}
}
}

\end{document}